\documentclass[10pt, conference, compsocconf]{IEEEtran}
\usepackage{cite}
\ifCLASSINFOpdf
  \usepackage[pdftex]{graphicx}
\else
\usepackage[dvips]{graphicx}
\fi
\usepackage[cmex10]{amsmath}
\usepackage{cite}
\usepackage{graphicx}
\usepackage{balance}  
\usepackage[tight,footnotesize]{subfigure}
\usepackage{clrscode3e}
\usepackage{float}
\usepackage{multirow}
\usepackage{wrapfig}
\usepackage{url}
\usepackage{booktabs}
\usepackage{array}
\usepackage{booktabs}
\usepackage{amssymb}
\usepackage{amsmath}
\usepackage{caption}
\newcommand{\superscript}[1]{\ensuremath{^{\textrm{#1}}}}
\newcommand{\alg}{\textsc{FS\superscript{3}}}

\def\qed{\rule{2mm}{2mm}}
\newcommand{\ra}[1]{\renewcommand{\arraystretch}{#1}}
\newtheorem{lemma}{Lemma}

\graphicspath{{}}

\begin{document}
\title{FS\raisebox{.6ex}{\LARGE 3}: A Sampling based method for top-k Frequent Subgraph Mining}

\author{\IEEEauthorblockN{Tanay Kumar Saha and Mohammad Al Hasan}
\IEEEauthorblockA{Dept. of Computer and Information Science\\ 
Indiana university---Purdue university Indianapolis, Indiana, IN-46202, USA\\
Email: \{tksaha,alhasan\} @cs.iupui.edu}
}

\maketitle
\begin{abstract}
Mining labeled subgraph is a popular research task in data mining because of
its potential application in many different scientific domains. All the
existing methods for this task explicitly or implicitly solve the subgraph
isomorphism task which is computationally expensive, so they suffer from the
lack of scalability problem when the graphs in the input database are large. In
this work, we propose \alg, which is a sampling based method. It mines a small
collection of subgraphs that are most frequent in the probabilistic sense.
\alg\ performs a Markov Chain Monte Carlo (MCMC) sampling over the space of a
fixed-size subgraphs such that the potentially frequent
subgraphs are sampled more often. Besides, \alg\ is equipped 
with an innovative queue manager. It stores the sampled subgraph 
in a finite queue over the course of mining in such
a manner that the top-$k$ positions in the queue contain the most frequent
subgraphs.  Our experiments on database of large graphs show that \alg\ is
efficient, and it obtains subgraphs that are the most frequent amongst the
subgraphs of a given size.
\end{abstract}

\maketitle
\section{Introduction}
Frequent Subgraph Mining (FSM) is an important research task. It has
applications in various disciplines, including
chemoinformatics~\cite{Deshpande.Kuramochi.ea:05},
bioinformatics~\cite{Hu.Yan.ea:05}, and social
sciences. The main usage of FSM is for finding
subgraph patterns that are frequent across a collection of graphs.  This task
is additionally useful in applications related to graph
classification~\cite{Jin.Calvin.ea:10}, and graph indexing~\cite{Yan.Yu.ea:04}.
However, existing algorithms for subgraph mining are not scalable to large
graphs that arise in social and biological domains~\cite{Chaoji.Hasan.ea:08*1}.
For instance, a typical protein-protein interaction (PPI) network contains a few
hundreds of proteins and a few thousands of known interactions. However the most
efficient of the existing FSM algorithms cannot mine frequent subgraphs in a
reasonable amount of time from a small database of PPI networks even with a
large support value~\cite{Chaoji.Hasan.ea:08*1} (also see,
Table~\ref{tab:tfinishps}). In this era of big data, we are collecting graphs
of even larger size, so an efficient algorithm for FSM is in huge demand.

Over the years, a good number of algorithms for FSM task were proposed; examples
include AGM~\cite{Inokuchi.Washio.ea:00},
FSG~\cite{Kuramochi.Karypis:04}, gSpan~\cite{Yan.Han:02}, and
Gaston~\cite{Nijssen.Kok:05}.  A common feature of these algorithms is that
they ensure completeness, i.e., they enumerate all the subgraphs that are
frequent under a user-defined minimum support.
For large graphs the subgraph space is too big to enumerate, so an algorithm that
traverses the entire space cannot finish in a feasible amount of time. In
fact, any exact method for frequent subgraph mining needs to solve numerous
Subgraph Isomorphism (SI)---a known ${\cal NP}$-complete problem, so the lack
of scalability of FSM is inherent within the problem definition.  One may
sacrifice the completeness and obtain a subset of frequent patterns as a
partial output by using one of the existing algorithms; however, because of artificial order
of enumeration imposed by the above mining algorithms, the patterns in the
partial output are not representative of the entire set of frequent patterns.

\begin{table}
\ra{1.2}
\small
\centering
\resizebox{\linewidth}{!}{%
\begin{tabular}{p{1.3cm} p{1.6cm}|p{1.2cm} p{1.4cm}| p{1.1cm} p{1.6cm}}
\multicolumn{6}{c}{\bf Dataset Statistics: \# graphs: 90, avg. \# vertices: 67,
avg. \# edges: 268}\\
\multicolumn{6}{c}{\bf ~~~~~~~~~~~~~~~~~~~~~~\# node labels: 20, \# edge labels: 3}\\[1em]
\toprule
\multicolumn{2}{c|} {\bf Time vs Max. subgraph size}&
\multicolumn{2}{c|} {\bf Time vs different minsup} &
\multicolumn{2}{c} {\bf Search Space vs subgraph size}\\
\multicolumn{2}{c|} {\bf (min-sup is fixed at 40\%)}&
\multicolumn{2}{c|} {\bf (Max-size is fixed at 8)} &
\multicolumn{2}{c} {}\\
\hline
Max-size & Time &  Support &Time &  Size  & Induced Subgraph	\tabularnewline
     &      &  (\%)    &     &        & Count 	\tabularnewline
\hline
\hline
8   &  6 minutes       &28 &1.1 hours    &6&26 millions\tabularnewline
9   & 2.8 hours        &22 &3.5 hours    &7&157 millions\tabularnewline
10  & $>$ 1.5 days     &17 &9 hours      &8&947 millions\tabularnewline
    &                  &11 &$>$16 hours  &9&5000 billions\tabularnewline
\bottomrule
\end{tabular}
}
\caption{\small Highlights of the lack of scalability of existing frequent subgraph mining 
methods while mining the PS dataset. Time indicates the running time of the fastest version of 
Gaston~\cite{Nijssen.Kok:05}}
\vspace{-0.2in}
\label{tab:tfinishps}
\end{table}

FSM's lack of scalability is well documented in many of the earlier
works~\cite{Chaoji.Hasan.ea:08*1,Hasan.Zaki:09*2}, yet we provide some
quantitative evidences so that a reader can comprehend the enormity of the
challenges. For this we mine subgraphs from a protein structure (PS) dataset (see
Section~\ref{myexperiment} for details) that contains only 90 graphs, each
having 67 vertices and 268 edges, on average. First we use a 64-bit binary of
gSpan software~\footnote{\small gSpan is the most polular among the existing graph mining
methods. We use the Linux binary made available by the inventors:
\url{http://www.cs.ucsb.edu/~xyan/software/gSpan.htm}}. Using a large 40\%
support, the mining task could last only for a few minutes, after that the OS
aborted the gSpan process because by that time it had consumed more than 80\% of 128
GB memory of a server machine. We then attempted the identical mining task
using Gaston software~\cite{Nijssen.Kok:05}~\footnote{\small Gaston is the fastest graph mining
algorithm at present as verified by independent comparison,
see~\cite{Worlein.Meinl.ea:05}}, which kept running for more than 2 days. Then
we ran the same software with a restriction on the maximum size of the subgraphs to be
mined (only Gaston allows such an option), yet the mining task seems to be
insurmountable. Table~\ref{tab:tfinishps} shows more detailed postmortem of Gaston's lack of
scalability for the subgraph mining task on the PS dataset.


To cope with the scalability problem, in recent years researchers have proposed
alternative paradigms of frequent subgraph mining, which are neither complete,
nor enumerative. Some of these works find frequent patterns considering their
subsequent application in knowledge discovery tasks. For example, there are
methods~\cite{Yan.Cheng.ea:08,Thoma.Cheng.ea:09} that directly mine frequent
subgraphs for using them as features for graph classification.  Another family
of works~\cite{Hasan.Zaki:09*2,Hasan.Zaki:09} perform MCMC random walk over the space
of {\em frequent} patterns and sample only a subset of all the frequent subgraphs. 
However, the above sampling based methods also solve numerous SI task for ensuring that the random walk traverses
only over the frequent patterns, so they are also not scalable when
the input graphs are large. 

There also exist some methods that find a subset of frequent subgraphs, such
as, frequent induced subgraphs (AcGM~\cite{Inokuchi.Washio.ea:02}), maximal
frequent subgraphs (SPIN~\cite{Huan.Wang.ea:2004},
MARGIN~\cite{Thomas.Valuuri.ea:2006}), or closed frequent subgraphs
(CloseGraph~\cite{Yan.Han:03}). In each of these cases, since the objective is
to mine a specific subset of frequent subgraphs, effective pruning strategies
can be exploited, which, sometimes, offer noticeable speed-up over traditional
frequent subgraph mining.  Nevertheless pruning typically offers a constant
factor speed-up, which is not much beneficial while mining large input graphs.
Also, like traditional subgraph mining all these methods perform numerous SI
tasks for ensuring the minimum support threshold, so they also are not scalable.
We ran both AcGM, and SPIN on the PS dataset; for a 10\% support both 
methods run for a while, but the mining task was aborted by the OS after the
software consumed more than 100 GB of memory.

Scalable subgraph mining is achievable if the database contains graphs from 
a restricted class for which the SI task is tractable (polynomial).
Some recent works on subgraph mining actually explored this option. Examples
include mining outerplaner graphs~\cite{Horvath:Ramon.ea:06}, or mining graphs
with bounded treewidth~\cite{Horvath.Tam.ea:2010}, or graphs where each of
the vertices have a distinct label~\cite{Vijayalakshmi.Nadarajan.ea:11}.
However, except chemical
graphs, for which the treewidth value is around $3$, general graphs from other
domains rarely adhere to such restrictions. The good mining performance on
treelike graph is probably the reason that the existing methods only use chemical
graphs for presenting their experiment results~\footnote{\small DTP dataset \\(available from
\url{http://dtp.nci.nih.gov/branches/dscb/repo_open.html}) is the most
popular graph mining dataset, which is mostly tree with an average vertex and edge size of 31 and 34
respectively.}. For general graphs the only viable option is to discard the
SI test altogether.  
A recent work, called GAIA~\cite{Jin.Calvin.ea:10} uses
this idea; however, the
scope of GAIA is limited for mining only discriminatory subgraphs that are good
for graph classification, so it is not applicable for mining frequent
subgraphs.

For frequent subgraph mining task, discarding SI test is possible, only if we
relax the minimum support constraint such that the returned subgraphs are
likely to be frequent, but they do not necessarily satisfy a user-defined
minimum support requirement. This seems to be an over-simplification which
evades the main purpose of frequent pattern mining---after all, in pattern
mining, the minimum support constraint is the threshold that decides which of
the candidate patterns are frequent and which are not. However, in practice,
the minimum support constraint has small significance, because a user seldom
knows what is the right value of minimum support parameter to find the best
patterns for her anticipated use~\cite{Keogh.Lonardi.ea:04}.  Further, it is a
hard-constraint which can discard a supposedly good pattern that narrowly
misses the support threshold.  An alternative to minimum support constraint can
be a size constraint, in which a user provides a size for the pattern that she
is looking for; in the context of subgraph mining, the size can be the number
of vertices (or edges) that a pattern should have. The argument in favor of
this choice is that it is easier for an analyst to define a size constraint
than defining a minimum support constraint using his domain knowledge---a size
constraint can be equal to the size of a meaningful sub-unit in the input
graph. For instance, if the input graph is a social network, a size constraint
can be equal to the size of a typical community in that network.

In this work, we propose a method for frequent subgraph mining, called \alg,
that is based on sampling of subgraphs of a fixed size~\footnote{\small The name \alg\
should be read as {\em F-S-Cube}, which is a compressed representation of the
4-gram composed of the bold letters in {\bf F}ixed {\bf S}ize {\bf S}ubgraph
{\bf S}ampler.}.  Given a graph database ${\cal G}$, and a size value $p$,
\alg\ samples subgraphs of size-$p$ from the database graphs using a 2-stage
sampling. In the first stage of a sampling iteration, \alg\ chooses one of the database
graphs (say, $G_i$) uniformly, and in the second stage it chooses a size-$p$ 
subgraph of $g$ using MCMC. The
sampling distribution of the second stage is biased such that it over-samples
the graphs that are likely to be frequent over the entire database ${\cal
G}$. \alg\ runs the above sampling process for many times, and uses an
innovative priority queue to hold a small set of most frequent subgraphs. The
unique feature of \alg\ is that unlike earlier works which are based on
sampling~\cite{Hasan.Zaki:09*2}, \alg\ does not
perform any SI test, so it is scalable to large graphs.
By choosing different values of $p$, a user can find a succinct set of frequent subgraphs
of different sizes. Also, as the number of samples increases, \alg's output progressively converges 
to the top-$k$ most frequent subgraphs of size $p$. So a user can run the
sampler as long as he wants to obtain more precise results.

\noindent We claim the following contributions in this work:\\

\noindent $\bullet$~~~We propose \alg, a sampling based method for mining top-$k$ frequent
subgraphs of a given size, $p$.  \alg\ is scalable to large graphs, because it
does not perform the costly subgraph isomorphism test.

\noindent $\bullet$~~~We design several innovative queue  mechanisms to hold the top-$k$
frequent subgraphs as the sampling proceeds.

\noindent $\bullet$~~~We perform an extensive set of experiments and analyze the effect of every control
parameter that we have used to validate the effectiveness and efficiency of \alg.


\section{Background}\label{bck}
\subsection{Graph, Induced Subgraph, Frequent Subgraph Mining}

Let $G(V,E)$ be a \textit{graph}, where $V$ is the set of vertices and $E$ is the
set of edges. Each edge $e \in E$ is denoted by a pair of vertices $(v_{i},
v_{j})$ where, $v_{i},v_{j} \in V$. A graph without self-loop or multi edge
is a simple graph. In this work, we consider simple, connected, and undirected
graphs. A \textit{labeled graph} $G(V,E,L, l)$ is a graph for which the
vertices and the edges have labels that are assigned by a labeling function, $l:V \cup E
\rightarrow L $ where $L$ is a set of labels. 

A graph $G' = (V',E')$ is a {\textit subgraph} of $G$ (denoted as $G' \subseteq G$) if $V'
\subseteq V$ and $E' \subseteq E$.  A graph $G' = (V',E')$ is a {\textit
vertex-induced subgraph} of $G$ if $G'$ is a subgraph of $G$, and for any pair
of vertices $v_a, v_b \in V'$, $(v_a, v_b) \in E'$ if and only if $(v_a, v_b)
\in E$. In other words, a \textit{vertex-induced} subgraph of $G$ is a graph $G'$
consisting of a subset
of $G$'s vertices together with all the edges of $G$ whose both endpoints are
in this subset. In this paper, we have used the phrase \textit{induced
subgraph} for abbreviating the phrase vertex-induced subgraph. 
If $G'$ is a (induced or non-induced) subgraph of $G$ and $|V'|=p$, we call $G'$ a $p$-subgraph
of $G$.


Let, ${\cal G} =\{G_1, G_2, \ldots, G_n \}$ be a graph database, 
where each $G_i \in {\cal G}, \forall i=\{1 \ldots n\}$ represents a labeled, undirected
and connected graph. {\bf t}$(g) = \{G_i : g \subseteq G_i \in {\cal
G}\}, \forall i = \{1 \ldots n \}$, is the {\em support-set} of the graph $g$.
This set contains all the graphs in ${\cal G}$ that have a subgraph isomorphic
to $g$. The cardinality of the {\em support-set} is called the {\em support} of
$g$. $g$ is called frequent if $support \geq \pi^{\textbf {min}}$, where
$\pi^{\textbf {min}}$ is predefined/user-specified \textit{minimum support
(minsup)} threshold. Given the graph database ${\cal G}$, and minimum support
$\pi^{\textbf {min}}$, the task of a frequent subgraph mining algorithm is to
obtain the set of frequent subgraphs (represented by ${\cal F}$).
While computing support, if an FSM algorithm enforces induced subgraph isomorphism,
it obtains the set of frequent induced subgraphs (represented by ${\cal F}_I$).
It is easy to see that ${\cal F} \subseteq {\cal F}_I$.\\

\subsection{Markov chains, and Metropolis-Hastings (MH) Method}\label{sec:mh}
The main goal of the Metropolis-Hastings algorithm is to draw
samples from some distribution $\pi(x)$, called the \textit{target
distribution}. 
where, $\pi(x) = f(x)/K$; here $K$ is a normalizing constant which may not be known and difficult to compute. 
It can be used
together with a random walk to perform MCMC sampling.
For this, the MH algorithm draws a sequence of samples from the
target distribution as follows: (1) It picks an initial state (say, $x$) satisfying $f(x) > 0$;
(2) From current state $x$, it samples a neighboring point $y$ using a
distribution $q(x,y)$, referred as \textit{proposal distribution}; (3) Then, it calculates the \textit{acceptance probability} given in Equation~\ref{eq:AP}, 
and accepts the proposal move to $y$ with probability $\alpha(x,y)$.  The
process continues until the Markov chain reaches to a stationary distribution.
In this work we used MH algorithm for sampling a size-$p$ subgraph from 
the database graphs.
\begin{equation}\label{eq:AP} \alpha(x,y)=\min
\Bigg(\frac{\pi(y)q(y,x)}{\pi(x)q(x,y)},1\Bigg)=\min
\Bigg(\frac{f(y)q(y,x)}{f(x)q(x,y)},1\Bigg) 
\end{equation}

\section{Problem Formulation and Solution Approach}\label{probform}
Our objective is to obtain a small collection of frequent subgraph patterns
from a database of large input graphs. For this, we aim to design a subgraph
mining method that does not perform the costly subgraph isomorphism (SI) test.
Without SI test, the exact support values of a (sub)graph in the database graphs are
impossible to obtain. So, we deviate from the traditional definition of {\em
frequent} that is used in the FSM literature, rather we call a graph frequent
if its {\em expected-support} (defined in the next paragraph) is comparably
higher than that of other same-sized graphs.  When a graph grows larger, its
support-set naturally shrinks, so keeping the size as an invariant makes sense,
otherwise the output set of our method will be filled with small patterns
(one-edge or two-edge) that have the highest support among all the frequent
patterns. However, note that the size is only a parameter, not a constraint;
i.e., a user can always run different mining sessions with different size
values as she desires.  A formal description of our research task is as
below: Given a graph database ${\cal G} = \{G_i: 1\le i\le n\}$ and a user-defined
size value $p$ it returns a list of top-$k$ frequent patterns, where {\em
frequent} is understood probabilistically. 

Our solution to this task is a sampling method, called \alg---a sampling
iteration of \alg\ samples a random size-$p$ subgraph (induced or non-induced
depending on the user requirement) $g$ from one of
the database graphs (say $G_i$), the later chosen uniformly. We call $g$
frequent, if an identical copy of it is sampled from many of the input graphs
in different sampling iterations of \alg.  In a sampling session, the number of
distinct input graphs from which $g$ is sampled is called its {\em
expected-support} and is denoted as $support_a(g)$.  Clearly actual support of
$g$ ($support(g)$) is an upper bound of the expected support of $g$
($support_a(g)$); generally speaking, these two variables are
positively correlated, so we use expected-support as a proxy of real support,
and thus \alg\ returns those $p$-subgraphs that are among the top-$k$ in terms
of {\em expected-support}. 

There are several challenges in the above solution approach.  First, when the
input graphs in ${\cal G}$ are large, for a typical $p$-value, the number of
possible $p$-subgraphs of $G_i$ is in the order of millions (or even billions,
see Table~\ref{tab:tfinishps}), so if we sample a $p$-subgraph from $G_i$
uniformly out of all $p$-subgraphs of $G_i$, the chance that we will sample a
frequent $p$-subgraph is infinitesimally small. Moreover, 
we do not know how many $p$-subgraphs exist for each of the input graphs
in ${\cal G}$, so a direct sampling method is impossible to obtain. To cope
with these challenges, \alg\ invents a novel MCMC sampling
which performs a random walk over the space of $p$-subgraphs of the graph
$G_i$; in this sampling, the desired distribution is non-uniform, which biases
the walk to choose $p$-subgraphs that are potentially frequent. Besides the
above, another challenge of our solution approach is that we do not have
unlimited memory, so during the sampling process,  we can store only a limited
number of sampled subgraphs in a priority queue; when the queue gets full, we
have to identify which of the  sampled subgraphs we will continue to maintain
in the queue. \alg\ solves this with a collection of novel queue management
mechanisms.

\section{Method}\label{sec:method}
\alg\ has two main components. A $p$-subgraph sampler, and a queue manager.
The first component samples a $p$-subgraph using MCMC sampling from a database
graph, $G_i$, later chosen uniformly. The second component maintains a
priority queue of top-$k$ frequent subgraphs of the input database ${\cal G}$.
We discuss each of the components in the following subsections.

\subsection{MCMC sampling of a $p$-subgraph from a database graph} 
The sample space of MCMC walk of \alg\ is the set of $p$-subgraphs of a database graph $G_i$.
At any given time, the random walk of \alg\ visits one of the $p$-subgraphs of
$G_i$. It then populates all of its neighboring $p$-subgraphs and
(probabilistically) chooses one from them as its next state using
MH algorithm. Below, we discuss the setup of MCMC sampling, including
target distribution, and state transition.\\

\noindent{\bf Target Distribution}: The target distribution of the MCMC walk of
\alg\ is biased so that the $p$-subgraphs that are likely to be frequent are
sampled more often. Formally, this distribution is a scoring function $f: \Omega
\rightarrow \mathbb{R}_+$; $f$ maps each graph in ${\Omega}$ (set of all
$p$-subgraphs) to a positive real number such that the higher the support of a
graph, the higher its score.  For efficiency sake, we want the scoring function
$f$ to be locally computable, and computationally light.  
It is not easy to find such a distribution up-front, because the support
information of a $p$-subgraph is not available until we discover that graph;
even if we have discovered the graph, and its partial
support is available to us, we cannot use that partial support information in
the target distribution, because if we do so it will bias the walk towards some
patterns that have already been discovered, but they may not be amongst the
most frequent ones. Also remember, \alg\ excludes the option of finding actual
support of a $p$-subgraph, because its goal is to avoid subgraph isomorphism tests
altogether.

\begin{figure}[ht]
\begin{minipage}[b]{0.225\textwidth}
\centering
\includegraphics[width=1.4in]{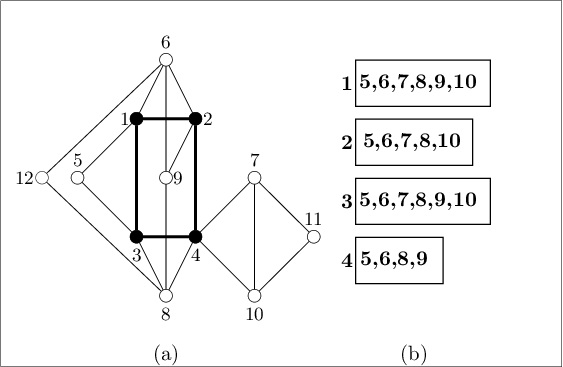}
\caption{(a) A database graph $G_i$ with the current state of \alg's random walk (b) 
Neighborhood information of the current state (1,2,3,4)}
\label{fig:neighbor}
\end{minipage}
\hspace{0.10in}
\begin{minipage}[b]{0.225\textwidth}
\centering
\includegraphics[width=1.4in]{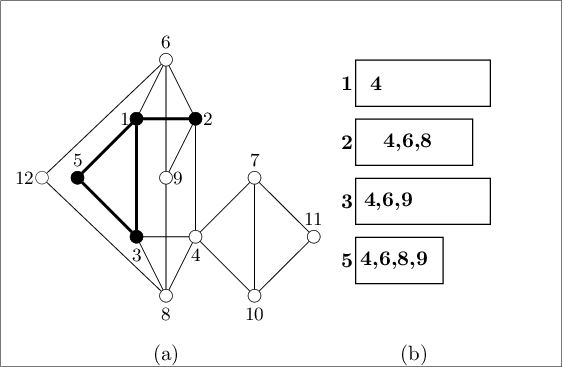}
\caption{(a) The state of random walk on $G_i$ (Figure~\ref{fig:neighbor}) after one transition 
(b) Updated Neighborhood information }
\label{fig:transition}
\end{minipage}
\end{figure}

In \alg, we have used two kinds of scoring functions: $s_1$ and $s_2$. For a
subgraph $g$, $s_1 (g)$ is the average of the (actual) {\em support} of the
constituting edges of $g$. Mathematically, 
$s_1(g) = \frac{1}{|E(g)|}\sum_{e \in E(g)}{support(e)}$.
$s_2(g)$ is  the cardinality of the
intersection set generated by intersecting the {\em support-set} of each of the
constituting edges of $g$, i.e.,
$s_2(g) = \left|\bigcap_{e \in E(g)}{support(e)}\right|$.
The intuition behind these choices is that if
$g$ is frequent, all its edges are frequent, so its score $s_1(g)$ is high, same is
true for $s_2(g)$. The reverse is not necessarily true, i.e., there
can be a graph, for which the average support  or the set intersection count of
its edge-set is high, but the graph is infrequent, so the above scoring functions may
sample a few false positive (however, no false negative) patterns. Nevertheless,
in real-life graphs the actual support
of a subgraph is significantly correlated with its $s_1$ and $s_2$ score, which we
will show in the experiment section.
Besides, when the sampling process discovers a  $p$-subgraph, 
its scores can be computed instantly from the {\em support-set}
of its edges---later can be obtained cheaply during
the initial read of the database graphs.\\


%

\begin{lemma} \label{score-lemma}
$s_1(g) \ge support(g)$ and $s_2(g) \ge support(g)$
\end{lemma}

\begin{proof}
See appendix.
\end {proof}

\noindent {\bf State Transition:} \alg's MCMC walk changes state by walking 
from one $p$-subgraph (say $g$) to a neighboring $p$-subgraph. In our neighborhood definition, 
for a $p$-subgraph
all other $p$-subgraphs that have $p-1$ vertices in common
are its neighbor subgraph/state. To obtain a neighbor subgraph of $g$, \alg\
simply replaces one of the existing vertices of $g$ with another vertex which
is not part of $g$ but is adjacent to one of $g$'s vertices. Also, note that
in $g$, if \alg\ includes all the edges of $G_i$ that are induced by the set
of the selected vertices, the sampled subgraph of \alg\ is always a connected
induced subgraph of the database graphs. On the other hand, if it does not
enforce this restriction, the sampled subgraph is a non-induced subgraph. 
Another important fact is that the neighborhood
relation that is defined above is symmetric, which is important in MCMC walk for maintaining 
the detailed balance equation~\cite{Rubinstein:08}.\\

\noindent {\bf Example:} Suppose \alg\ is sampling $4$-subgraphs from the graph
$G_i$ shown in Figure~\ref{fig:neighbor}(a) using MCMC sampling. Let, at any
given time the $4$-subgraph $\langle 1, 2, 3, 4\rangle$ (shown in bold lines)
is the existing state of this walk. Figure~\ref{fig:neighbor}(b) lists all the
neighbor states of this state. In this figure, the box labeled by $x$ contains
all the vertices that can be used as a replacement of vertex $x$ to get a
neighbor. For example, one of the neighbor states of the above state is
$\langle 1, 2, 3, 5\rangle$, which can be obtained by replacing the vertex $4$
by the vertex $5$. If the \alg's random walk transition chooses to go to the
neighbor state $\langle 1, 2, 3, 5\rangle$, it can do it simply by adding the
vertex 5 (a vertex in the box labeled by 4) and deleting the vertex $4$.
While adding vertex 5, it adds both the induced edges
$(1,5)$ and $(3,5)$ for obtaining an induced subgraph, but adding a random
subset of the set of induced edges would have sampled a $p$-subgraph which is not necessarily
induced. The updated state of the random walk along with the updated neighbor-list is shown
in Figure~\ref{fig:transition}.~\qed \\

\noindent {\bf Proposal Distribution}: As discussed in Section~\ref{sec:mh},
for applying MH algorithm, we also need to decide on a proposal distribution,
$q$. For \alg's random walk the proposal distribution is uniform, i.e., in the
proposal step \alg\ chooses one of $g$'s neighbors uniformly. If a $p$-subgraph $g$  
has $d_g$ neighbors, and $h$ is one of them, using proposal
distribution, the probability of choosing $h$ from $g$ is $q(g, h) = 1/d_g$.\\

\begin{figure}
\centering
\scalebox{0.65}{
\fbox{
\begin{minipage}[t]{0.65\textwidth}
\begin{codebox}
\Procname{\proc{SampleIndSubGraph}($G_i,p$)}
\li $x\gets$ State saved at $G_i$ 
\li $d_x\gets$ Neighbor-count of $x$
\li $a\_sup_x\gets$ score of graph $x$
\li \While (a neighbor state $y$ is not found) 	
	\Do
\li $y\gets$ a random neighbor of $x$
\li $d_y\gets$ Possible neighbor of $y$
\li $a\_sup_y\gets$ score of graph $y$
\li $accp\_val\gets$ ${(d_x *a\_sup_y)}/{(d_y*a\_sup_x)}$
\li $accp\_probablility\gets$ $min(1,accp\_val)$
\li \If $uniform(0,1) \leq  accp\_probability$
	\Then
\li	\Return $y$
\End
\End
\end{codebox}
\end{minipage}
}}
\caption{\proc{SampleIndSubGraph}\   Pseudocode}
\label{fig:sub}
\vspace{-0.2in}
\end{figure}

In Figure~\ref{fig:sub} we show the MH subroutine that samples a $p$-subgraph
from a database graph $G_i$.  In Line 1, it obtains the $p$-subgraph, $x$ (a
state of the Markov chain) that was saved during the last sampling from $G_i$
in one of the previous iterations. If the saved state is empty (happens only if
it is the first graph sampled from $G_i$), it simply obtains one of the
$p$-subgraphs by growing from a random edge of $G_i$ and returns it. In Line 2,
it populates the neighbors of $x$ and returns the neighbor-count. In Line 3, it
computes the score of the graph $x$ based on the chosen scoring function ($s_1$ or $s_2$).
It then chooses $y$  uniformly from all the neighbors of $x$, populates
the neighbors of $y$ and computes $y$'s score (Line 5-7). Considering
the chosen scoring function as the desired target distribution, it computes the
acceptance probability of the transition from $x$ to $y$ using
Equation~\ref{eq:AP}. The while loop (Line 4-11) continues until a valid
next state (a neighboring $p$-subgraph) is found. It then returns the newly sampled
subgraph $y$.\\


\begin{lemma} \alg's random walk is ergodic.
\end{lemma}
\begin{proof}
See appendix.
\end{proof}
\begin{lemma}
The random walk of \alg\ achieves the target probability distribution, which is proportional
to the chosen scoring function ($s_i$)
\end{lemma}

\begin{proof}
See appendix.
\end{proof}

\subsection{Queue Manager}\label{qevict}

\alg\ runs the $p$-subgraph sampler for a large number of iterations so that in
these iterations, the most frequent patterns have a chance to be sampled a
number of times that is proportional to its support. Since the number of
possible $p$-subgraphs in a database of large graphs can be very large,
it may not be feasible to store all of them in the main memory.  So \alg\ stores
only a finite number of {\em best} graphs in a priority queue. The queue
manager component of \alg\ implements the policy of this priority queue (PQ).

For a graph, $g$, stored in the PQ,  the queue manager stores four pieces of
information regarding the graph: (1) the canonical label~\footnote{\small canonical
label is a string represent of a graph which is unique over all isomorphisms of
that graph; for our work we use min-dfs canonical code which is discussed
in~\cite{Yan.Han:02}} of $g$; (2) the {\em expected-support} value ($support_a(g)$) 
at that instance along with the support-list; (3) the score of $g$,
i.e. $s_1(g)$ or $s_2(g)$ depending on which of the target distribution is
used; and (4) the time (iteration counter is used as time variable) when the
$support_a(g)$ was last incremented.  The canonical label is used to uniquely
identify a graph in PQ to overcome the fact that different sampling iterations may return different
isomorphic forms of the same graph. The other pieces of information are used to
implement the policy of the PQ. 

{\bf Queue Eviction Strategy} If the new sample is an existing graph
in PQ, no eviction is necessary. We simply insert the id of the corresponding
database graph (from where the sample was obtained) into the support-list of
the graph and update the time variable. In case the id already is present in the
support-list, nothing happens. On the
other hand, if the new sample is a graph that does not present in PQ and PQ is full, we
may choose to accommodate the new graph by evicting one of the graphs in the PQ,
if certain conditions are satisfied.

To expedite the eviction decision, we maintain a total order in the PQ using a
composite order criterion and the last graph in that total order is
possibly evicted. The order uses three variables in lexicographical
order: (1) expected-support (high to low); (2) score value, $s_1$ or $s_2$,
depending on which one is used as the target distribution of the MCMC sampling
(high to low); and (3) time (recent to old). Thus, the
graph with the least expected support occupies the last position in PQ.
However, if more than one graphs have the same value for the least
expected-support, the tie situation is resolved by placing the graph with the
smallest score value in the last position. Note that for \alg's sampling, tie
on expected-count is common as the search space is very large. If there is a
tie for the score value also, it is resolved by considering the graph with the
oldest update time. The intuition behind the above eviction mechanism is easy
to understand; The pattern in the last position has small expected-support
(first criterion), or small score, $s_1$ or $s_2$ (second criterion), or it is
not being sampled from different graphs for a long time (third criterion), which makes
it less likely to be frequent. 

However, \alg's queue manager does not simply evict the last element in PQ to
insert the newly sampled graph (say, $g$), rather it first confirms whether $g$
is a better replacement for the graph that would be evicted from the PQ. The
decision is made by using the following heuristic. If the average of the scores 
($s_1$ or $s_2$) of the graphs that
are at the tail (lower half) of the PQ is smaller than $s_1(g)$
(or $s_2(g)$), then $g$ is considered as a better replacement, and the last
graph in the sorted order is evicted. If the above condition does not satisfy,
graph $g$ is simply ignored, and the sampling continues. The biggest advantage
of this conditional eviction is that \alg\ does not generate the canonical code of 
$g$, if $g$ is an unpromising pattern. Since, canonical code generation is much
costlier than sampling, the time saved by avoiding the code generation can be
spent for performing many other sampling iterations.
For implementing the data structure of queue manager with the queue eviction
policies, \alg\ uses multi-index map data structure~\footnote{\small We used boost
multi-index container
(\url{http://www.boost.org/doc/libs/1_53_0/libs/multi_index/doc/index.html}) as our
data structure},
which sorts the graphs uniquely on the canonical label and non-uniquely on the
various criteria that we describe above.

\begin{figure}
\centering
\small
\vspace{0.0in}
\scalebox{0.65}{
\fbox{
\begin{minipage}[t]{0.65\textwidth}
\begin{codebox}
\Procname{\proc{\alg}$({\cal G}, p,mIter)$}
${\cal G}$:~Graph Database, $p$:~Size of the subgraph\\ 
$mIter$:~Number of samples\\
\li $iter \gets 0$, $Q \gets \emptyset$\\
\li \While $iter\le mIter$
    \Do 
\li  $iter = iter +1$
\li Select a graph $G \in {\cal G}$ uniformly
\li $h \gets \proc{SampleIndSubgraph}(G,p)$
\li \If $Q.full = true$ {\bf and}\\
        $~~~~~~~~~~~~~~~~~~h.score() < Q.lowerHalfAvgScore()$\\
\li         \Then {\bf continue}
    \End
    
\li $h.code \gets \proc{GenCanCode}(h)$ 
\li \If $h \in Q$
    \Then   
\li $prevSupport = h.idset.size()$
\li $h.idset = h.idset \cup G.id$
\li \If $h.idset.size() > prevSupport$
\li    \Then 	$h.insertTime= iter$
    \End
\li \Else
\li  \If $Q.full = true$
\li      \Then $Q.evictLast()$
     \End
\li      $h.idset = \{G.id\}$
\li	 $h.insertTime=iter$
\li	 $Q = Q \cup \{h\}$
    \End
    \End
\li \Return $Q$

\end{codebox}
\end{minipage}}}
\caption{\alg\ Pseudocode}\label{fig:alg} 
\vspace{-0.04in}
\end{figure}

\subsection{Theoretical Analysis of \alg}
Theoretical analysis of \alg\ is difficult as the distribution of $p$-subgraphs 
is different for different datasets. We perform a theoretical analysis using a uniform
distribution which is given in the Appendix. 

\subsection{\alg\ Pseudocode}
The entire pseudo-code of \alg\ is shown in Figure~\ref{fig:alg}. It samples a
$p$-subgraph ($h$) from a randomly selected database graph $G$ by calling {\sc
SampleIndSubGraph} routine. Line 7 ensures that the sampled
graph $h$ is ignored (and its canonical code is not generated) if its score is not
better than the average score of the lower-half graphs in the PQ. In subsequent lines,
If $h$ does not present in the priority queue PQ,
\alg\ saves the graph $h$ in the priority queue along with its support-list
which contains only $G.id$. On the other hand, if $h$ exists in the queue,
\alg\ updates its support list, and also updates its insert-time variable.  For
each graph $G \in {\cal G}$, the sampling process saves the latest visiting
graph (state), so that any later sampling from this graph starts from the saved
state. From this perspective, \alg\ runs $|{\cal G}|$ copy of MCMC samplers,
one for one of the input graphs in ${\cal G}$. 


\section{Experiments}\label{myexperiment}

We implement \alg\ as a C++ program, and perform a set of experiments for
evaluating its performance for mining frequent subgraphs of a given
size. We run all the experiments in a computer with 2.60GHz processor and 4GB
RAM running Linux operating system. 

\subsection{Datasets}
We use three datasets for our experiments. The first is a protein structure
dataset that we call PS.  In this dataset, each graph represents the structure
of a protein in the TIM (Triose Phosphate Isomerage) family. To construct a
graph from a protein structure, we treat each amino acid residue as a vertex
(labeled by letter code of the amino acids), and connect two vertices with an
edge if the Euclidean distance between the $C_\alpha$ atom of the corresponding
residues is at most $8$\AA. An edge also has a label of 1 or 2 based on whether
the distance is below or above $4$\AA. Frequent subgraphs in such a dataset are
common structure of the homologous proteins. The statistics of this dataset are
available in Table~\ref{tab:tfinishps}; the same table also shows that existing
graph mining methods are not able to mine subgraphs from this dataset. Our
second dataset is a synthetic dataset (we call it Syn) that we build using the generator
used in~\cite{Cheng.James:07} with parameters (ngraphs, size, nnodel, nedgel)
$=$(0.1, 250, 20, 5). The subgraph space of this dataset is even larger than
the PS dataset, and hence, it is more difficult to mine.
Our last dataset is called Mutagenicity II (we will call it Mutagen dataset for
abbreviation); it has been used in earlier works on graph
mining~\cite{Bringmann.Zimmermann.ea:06}. Note that, it contains mostly
chemical graph (avg.\ vertex count=14, avg.\ edge count=14), and existing graph
mining methods can mine this dataset easily. We use this dataset only for
comparing precision because ground truth of frequent subgraphs for this
graph is easy to obtain.

\begin{figure}[ht]
\begin{center}
\subfigure[\#$p = 7$]{
\label{fig:7-kendall-ps}
\includegraphics[width=0.22\textwidth] {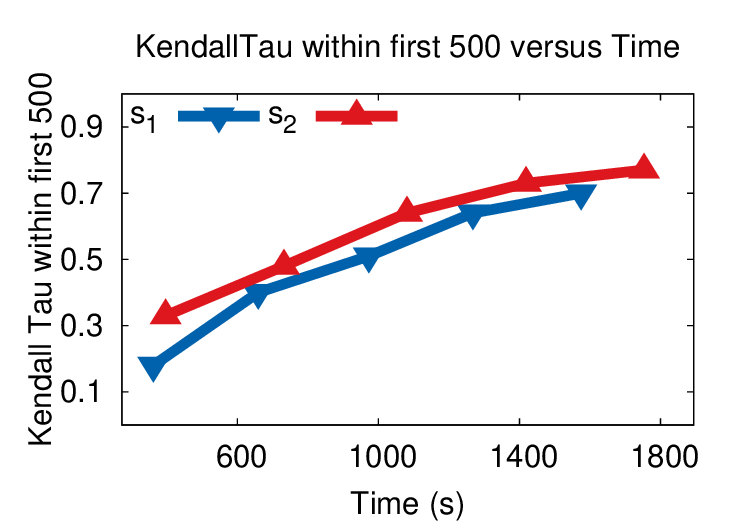}
}
\subfigure[\#$p = 7$]{
\label{fig:7-precision-ps}
\includegraphics[width=0.22\textwidth] {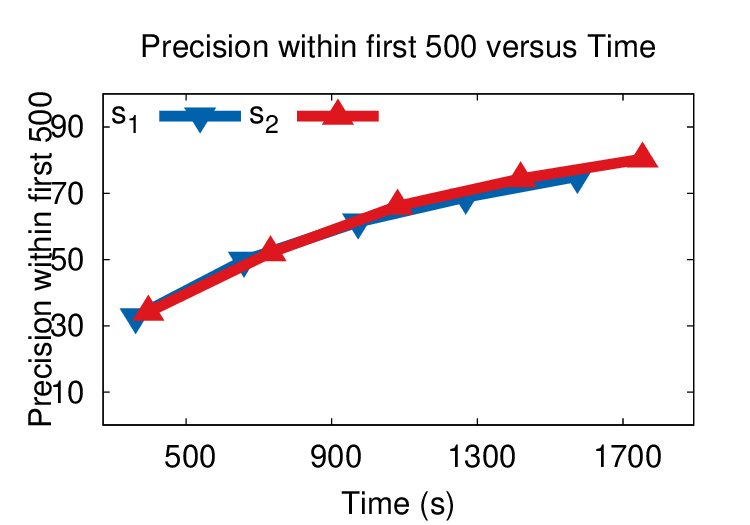}
}
\subfigure[\#$p = 8$ ]{
\label{fig:8-kendall-ps}
\includegraphics[width=0.22\textwidth] {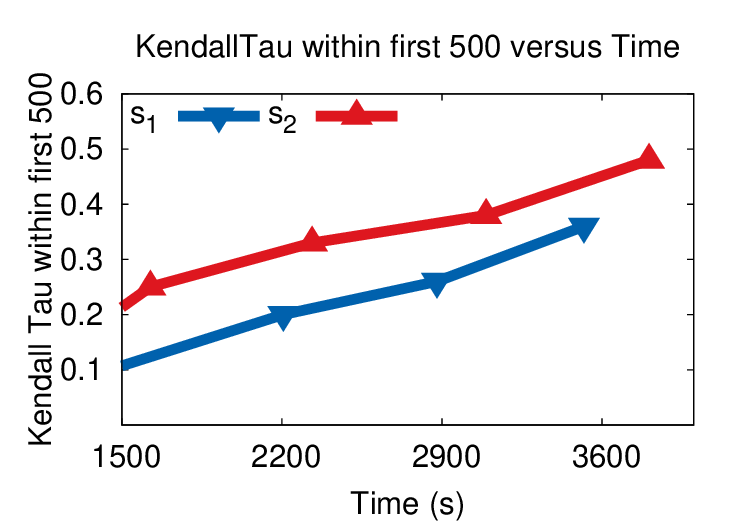}
}
\subfigure[\#$p = 8$]{
\label{fig:8-precision-ps}
\includegraphics[width=0.22\textwidth] {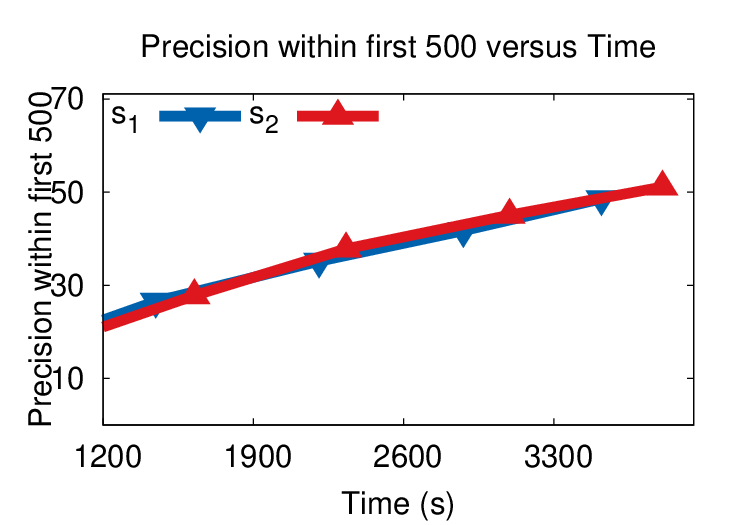}
}
\end{center}
\vspace{-0.2in}
\caption{Kendall Tau, Precsion within first 500 for PS Dataset}
\label{fig:ps-graphs}
\end{figure}

\begin{figure}[ht]\centering 
\subfigure[$p = 6$]{
\label{fig:6-convergence-ps}
\includegraphics[width=0.22\textwidth] {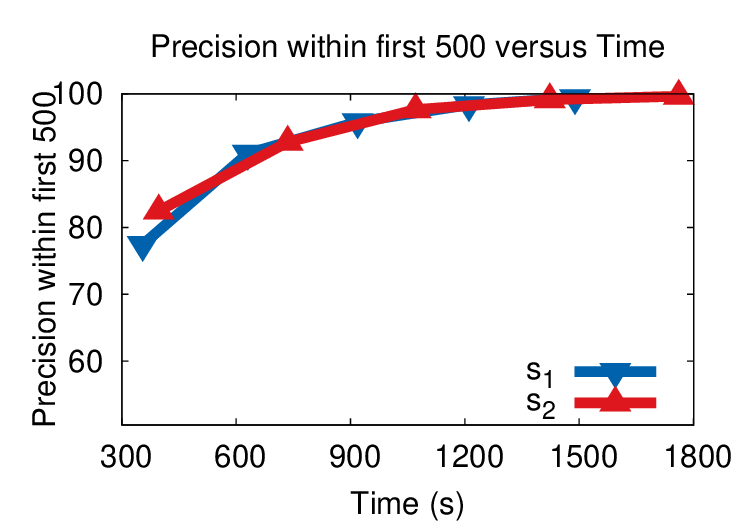}
}
\subfigure[$p = 7, 8$]{
\label{fig:7-convergence-ps}
\includegraphics[width=0.22\textwidth] {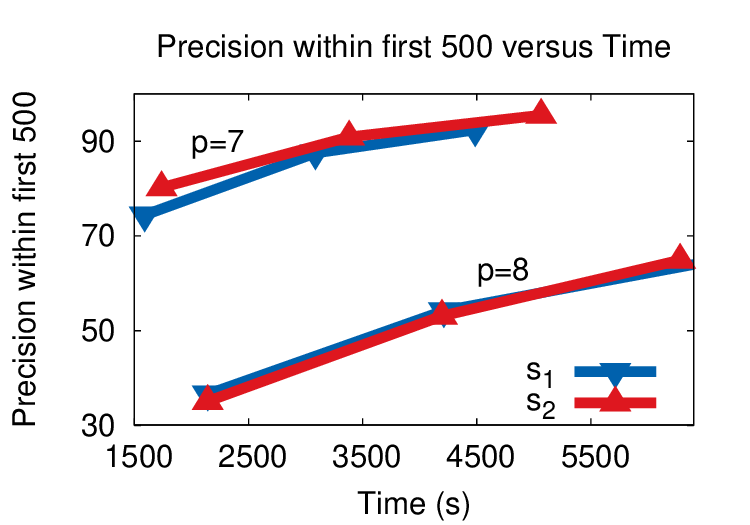}
}
\caption{Effect of increasing running time for \alg versus precision for PS Dataset}
\label{fig:effect-inc-time}
\end{figure}

\subsection{Experiment Setup}
\alg\ finds top-$k$ frequent subgraphs with high probability. So, we measure
the performance of \alg\ both from the execution time, and the quality of
results. To obtain the quality, we use two metrics, that are pr@500 (precision at 500),
and rank correlation metric, Tau-$b$. If ${\cal H}_a$ is the set of 500 most frequent
subgraphs of a given size obtained by \alg\, and ${\cal H}$ is the corresponding
true set of the same size based on actual support, the metric pr@500 is 
$\frac{\left|{\cal H} \cup {\cal H}_a \right|\times 100}{500}$; i.e, it finds the percentage
of graphs in ${\cal H}$ that are also presented in ${\cal H}_a$. The higher the value
of pr@500, the better the performance of \alg. Note that, for a graph dataset that has
one billion of subgraphs of a given size, sampling frequent graphs that
belong to set ${\cal H}$ is not easy. A dumb sampler has a pr@500 value equal to 500 
divided by one billion.

The metric, pr@500 only considers the presence or absence of a true positive
(actually frequent) graph in ${\cal H}_a$, but it does not consider the order
of graphs in ${\cal H}_a$ and the order of graphs in ${\cal H}$; in other words,
it does not check whether actual support and expected support (as obtained by
\alg) have positive correlation or not.  For this we use Tau-$b$ metric,
which is the rank correlation between
actual support and expected support of the objects in ${\cal H} \cup {\cal H}_a$.
Tau-$b$ varies between -1 and 1. A value
of 0 means no correlation, and the higher the value above 0, the better the
correlation. A strong correlation provides the evidence that \alg\ can indeed
rank the patterns in the order of their actual support.

For computing pr@500 and Tau-$b$, we need to know the true set of top 500
frequent patterns of a given size. This is difficult to obtain for PS and Syn
dataset, which we cannot mine with the existing methods.
To solve this problem, we
have used GTrieScanner~\cite{Ribeiro.Silva:10}; for an input graph GTrieScanner
dumps all of its $p$-subgraphs; by running this program for all the input
graphs in a graph database, and grouping those by the canonical-code of those
$p$-subgraphs, we compute the actual support value of all the $p$-subgraphs.
Such exhaustive enumeration of actual support was only possible for the Mutagen
dataset for all sizes, and for the PS and Syn datasets for size up to 8. For the
later two datasets, for size larger than 8, the size of the dump of
GTrieScanner exceeds more than 1 TB of physical space of a hard-disk, which is
impossible for us to post-process. Also note that, GTrieScanner generates only
the induced subgraphs, so for this comparison we run \alg\ for its induced
subgraph sampling setup.

Performance of \alg\ depends on the number of iterations, scoring function
used, size of the sampled patterns, and of-course the dataset. Also, choices of
these values affect the running time of an iteration.  So, when comparing
among different sampling scenarios of \alg\, we plot the performance metric
along the $y$-axis and the time along the $x$ axis, and use a smooth curve to
show the trend.  Since, our method is randomized, all performance metric values
are average of 10 distinct runs. We keep the priority queue size at 100K for
all our experiments, unless specified otherwise (memory footprint around 200
MB). Majority of our results are obtained by running experiments on the PS dataset.

\begin{figure}[ht]\centering 
\subfigure[Correlation with $s_1$]{
\label{fig:actualvsavg}
\includegraphics[width=0.22\textwidth] {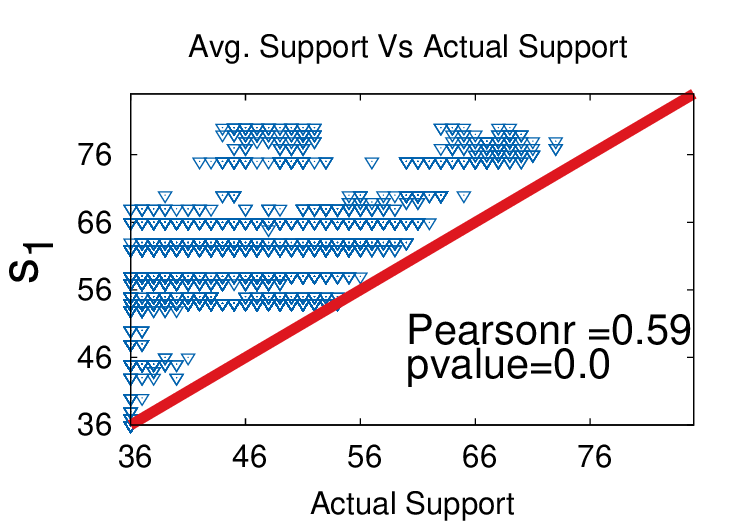}
}
\subfigure[Correlation with $s_2$]{
\label{fig:actualvssetinter}
\includegraphics[width=0.22\textwidth] {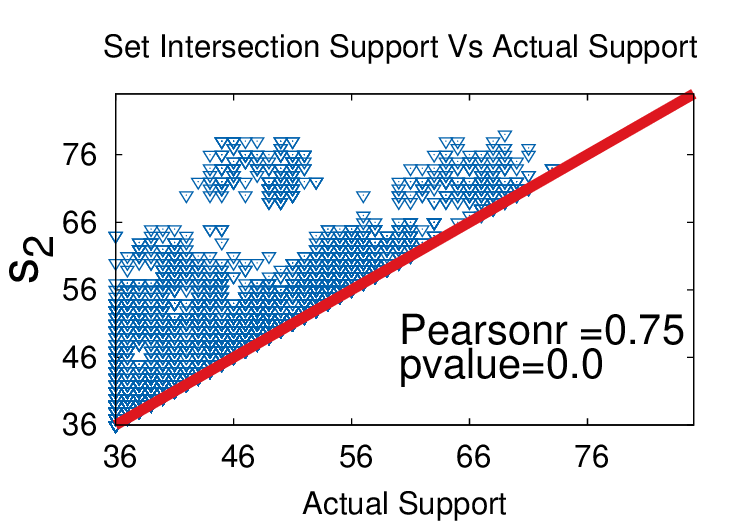}
}
\caption{Correlation between support and score of a pattern}
\label{fig:correlation}
\vspace{-0.2in}
\end{figure}

\subsection{Correlation between actual support and scores}\label{sec:corr}

In Figure~\ref{fig:actualvsavg} and \ref{fig:actualvssetinter}, 
we show the scatter plot between actual support vs $s_1$ value
(left plot) and $s_2$ value (right plot) of these patterns. 
This significant correlation between actual support and scores enables the \alg's MCMC
walk to be able to sample top-$k$ frequent patterns effectively.



\subsection{Performance of \alg\ for different sampling setups}\label{sec:samsetup}
In this experiment, we compare the performance of \alg\ using the scoring
function $s_1$ and $s_2$ on PS
dataset for size 7 and 8 (the true set (${\cal H}$) is known for these sizes).
Figure~\ref{fig:ps-graphs} shows the results; in the left, we show the results
(pr@500, and Tau-$b$ vs time) for size 7, and in the right for the size 8.
>From the figure, we see that for both the scores, with increasing number of
samples both pr@500, and Tau-$b$ metrics increase almost linearly.
Another observation from this figure is that the choice of score
($s_1$ or $s_2$) has small effect on the performance metric, specifically for
pr@500. For Tau-$b$, score $s_2$ performs slightly better than the score $s_1$.
This trend holds for other two datasets also. 

Now, we comment on the values of pr@500 and Tau-$b$ on these figures.
>From Figure~\ref{fig:ps-graphs}(d), we see that for size 8,
1500 seconds of running of \alg\ yields pr@500 value of 28\%, which increases to
50\% for 3700 seconds, i.e., within an hour of sampling time,
\alg\ finds 50\% of the most frequent graphs from a sampling space of 0.95
billions graphs (See Table~\ref{tab:tfinishps}).  Also note that the fastest
graph mining algorithm, Gaston, could not mine this dataset in 16 hours
of time, for 11\% support and the max-size of 8. Also, within an hour of
running, \alg's Tau-$b$ value reaches up to 0.42, which is a significant
correlation. Now, for size 7, the performance is understandably better than the size 8 (see 
figure~\ref{fig:ps-graphs}(a) and (b)), because its search space contains smaller
number of subgraphs---157 millions as reported in Table~\ref{tab:tfinishps}.

\begin{figure}[ht] \centering 
\subfigure[Syn, \#p = 6] {
\label{fig:pre-syn}
\includegraphics[width=0.22\textwidth]{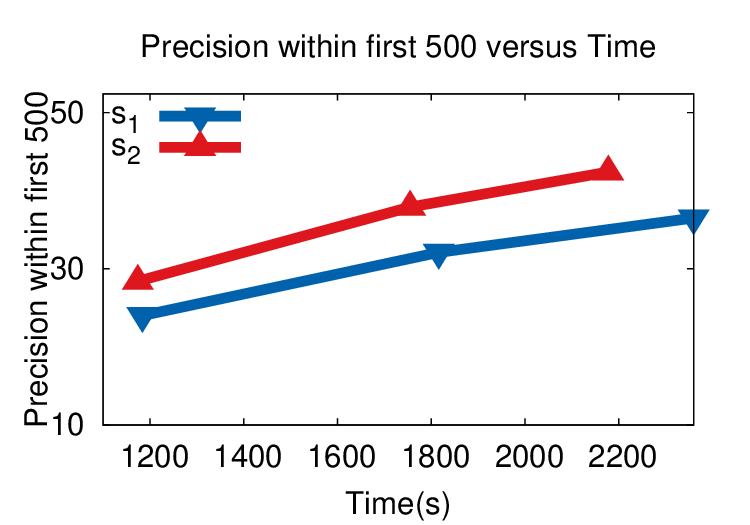} }
\subfigure[Mutagen, \#p = 8, 9, 10] { \label{fig:pre-muta}
\includegraphics[width=0.22\textwidth]{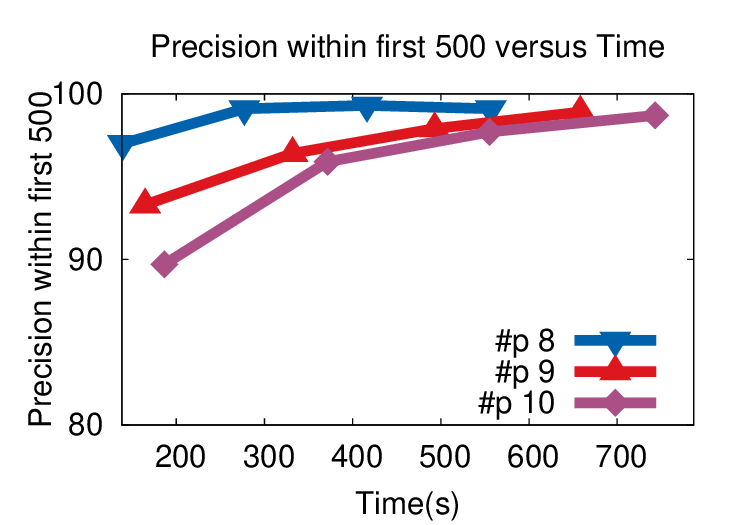} }
\caption{Precision for Synthetic and Mutagen Dataset} \label{fig:pre-syn-muta}
\vspace{-0.2in} 
\end{figure}

What happens if we run \alg\ for even more iterations? The performance keeps improving as we see in
Figure~\ref{fig:effect-inc-time}. By running the sampler for 20 minutes for
size 6, 1.4 hour for size 7, and 1.8 hour for size 8,  we obtain 99\%, 95\% and
65\% value for the pr@500. The linear trend of the curve for size 8 shows
that by running more time, the pr@500 can be improved even further.

We also run the above set of experiments for the other two datasets. 
In Figure~\ref{fig:pre-syn},
we show the results for Syn dataset for size 6, for which we
obtain pr@500 value of 42\% in around 35 minutes. The performance on this
dataset is poorer than the PS dataset, because search space in this
dataset is much larger than the PS dataset. We cannot show results
for higher size for this dataset as we could not generate the ground truth.
In Figure~\ref{fig:pre-muta}, we show the results for the Mutagen dataset,
which has the smallest subgraph space, so for sizes 8, 9, and 10 this
dataset achieves more than 90\% pr@500 within 10 minutes.

\begin{figure}[ht]
\centering
\subfigure[PS]{
\includegraphics[width=0.22\textwidth]{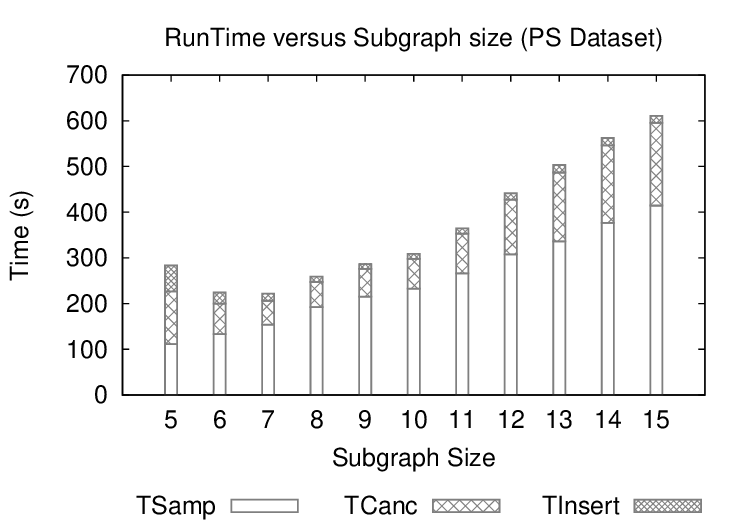}
\label{fig:runtime-ps}
}
\subfigure[Syn]{
\includegraphics[width=0.22\textwidth]{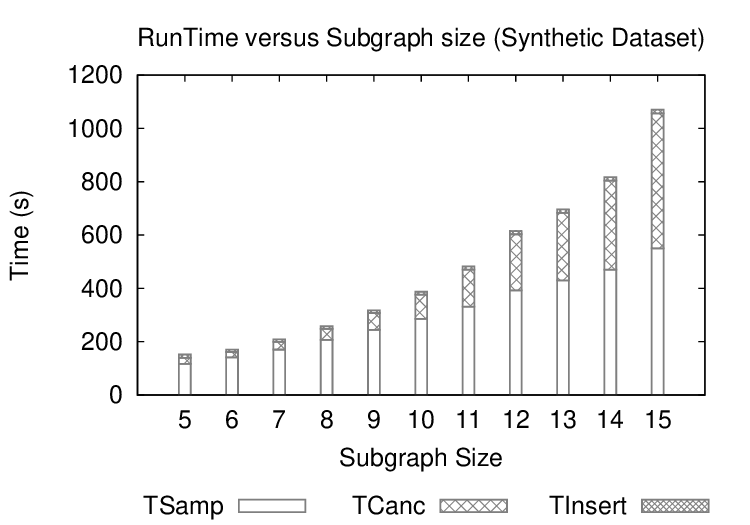}
\label{fig:runtime-syn}
}
\caption{Runtime performance of \alg\ for sampling subgraphs of different sizes}
\label{fig:runtime}
\vspace{-0.2in}
\end{figure}

\subsection{\alg's scalability with the size, $p$}
The execution time of \alg\ has three components: sampling time, canonical code
generation time, and queue insertion time. In this experiment, we check how
these times vary as we vary the desired size of the subgraphs to be sampled
($p$ value). For this, we use PS and Syn datasets, and use $s_2$ scoring function.
Figure~\ref{fig:runtime} shows the results. As we see in the
plots, the execution time increases almost linearly with the value of $p$ for
both the datasets.  Also, \alg\ spends the majority of its execution time for
sampling as it does not generate canonical code in
many of its iterations. Queue insertion time is negligible compared to sampling
and canonical code generation time.

\begin{figure}
\subfigure[Effect of Target Distribution]{
\includegraphics[width=0.22\textwidth]{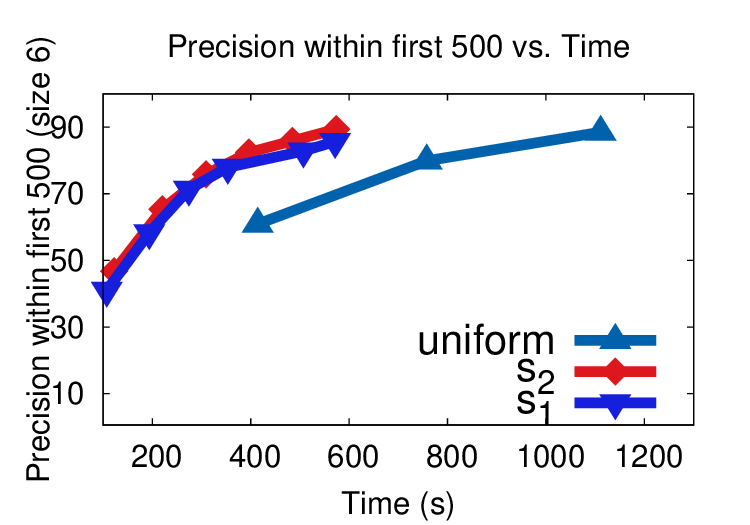}
\label{fig:comp}
}
\subfigure[Effect of Queue Size]{
\begin{minipage}{0.22\textwidth}
\vspace{-0.8in}
\centering
\bf
\scalebox{0.65}{
\begin{tabular}{c|c|c|c}
\hline
Queue&Precision&Kendal&Time\tabularnewline
Size&&&(s)\tabularnewline
(m)&&&\tabularnewline
\hline
\hline
\small
0.5&52.3&35.4&3997\tabularnewline
\hline
1.0&54.2&42.4&4211\tabularnewline
\hline
2.0&53.9&55.4&4645\tabularnewline
\hline
\hline
\end{tabular}
}
\vspace{0.2in}
\end{minipage}
\label{fig:equeuesize}
}
\caption{Effect of Queue Size and Target Distribution }
\label{effect-q-target}
\vspace{-0.3in}
\end{figure}

\subsection{Impact of target distribution and queue size} \label{sec:impact}
\alg's MCMC sampling uses $s_1$ or $s_2$ score to construct its target distribution.
In this experiment, we validate the impact of these choices by comparing their performances with a
case, where the target distribution is uniform, i.e., each of the $p$-subgraphs
of a database graph $G_i$ has equal likelihood to be visited, that is the score
of any $p$-subgraph is 1, a constant (let's call it uniform-\alg). For comparison,
we use the pr@500 metric. Figure~\ref{fig:comp} shows the result for  PS dataset for size 6.
It is clear from this figure that by
adopting $s_1(g)$ or $s_2(g)$ as the target distribution, we achieve higher pr@500 at a
faster rate. For example, within 7 minutes of sampling, the pr@500 score 
of uniform-\alg\ is around 55\%; on the other hand, for the same time, 
the pr@500 score is around 85\% for both $s_1(G)$ and $s_2(G)$.

For all our experiments we kept the priority queue size fixed to 100K.  If we
increase the queue size, the memory footprint of the algorithm will increase,
but the method will be more accurate, as it will be able to store a large
number of potential frequent graphs that may turn out to be frequent at a later
time. The improvement is more prominent for the Tau-$b$ metric than the pr@500
metric as shown in Figure~\ref{fig:equeuesize} for PS dataset and subgraph size
8.

\subsection{Impact of k on \alg}
\begin{figure}[ht]
\centering
\subfigure[Precision]{
\includegraphics[ width=0.22\textwidth]{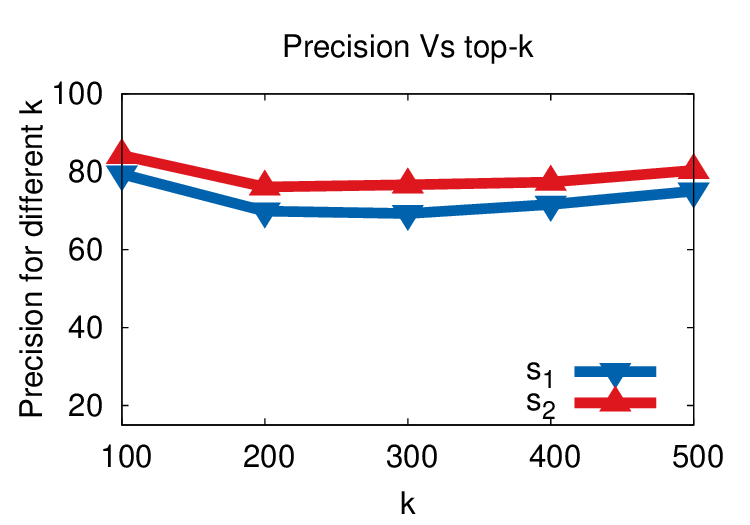}
\label{fig:impact-k-precision}}
\subfigure[Kendall Tau]{
\includegraphics[width=0.22\textwidth]{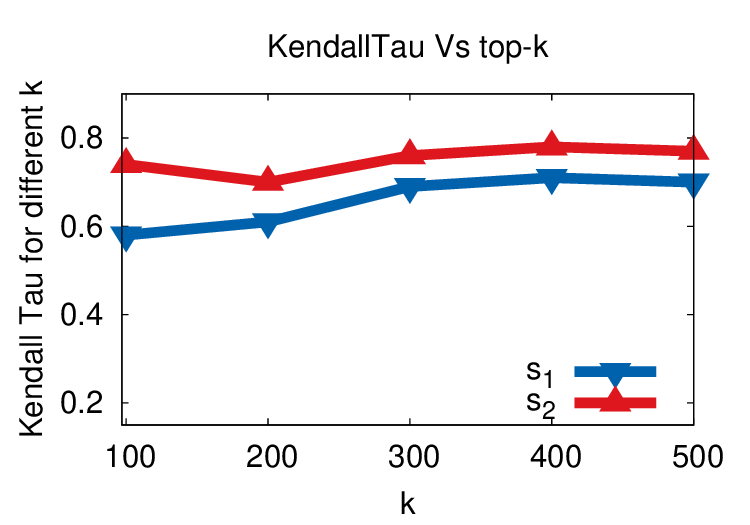}
\label{fig:impact-k-kendall}
}
\caption{Performance of \alg\ for different $k$}
\label{fig:impact-k}
\vspace{-0.2in}
\end{figure}

We also study the performance of \alg\ for different choices of $k$ value.  For
this experiment, we use PS dataset, $p$=7. Figure~\ref{fig:impact-k} shows the corresponding 
result. In Figure~\ref{fig:impact-k-precision}, we plot the Pr@k values and in
Figure~\ref{fig:impact-k-kendall}, we plot the Kendall Tau values for
different $k$'s between $100$ and $500$. We calculate both the statistics by taking
the average of $10$ independent runs. As we can see, for the entire range of
$k$ values, the performance remains almost constant.

\begin{figure}[t!]
\centering
\includegraphics[width=0.55\linewidth]{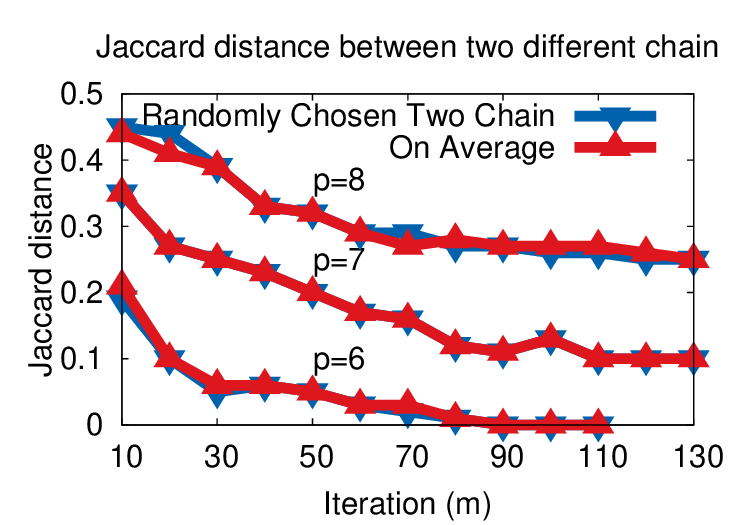}
\label{fig:jaccard-dist}
\caption{Jaccard Distance vs Iteration}
\label{fig:jaccard}
\vspace {-0.3in}
\end{figure}

\subsection{Choosing iteration counts}

We design a more sophisticated
stopping criteria using Gelman-Rubin Diagnostics~\cite{Gelman.Rubin:1992}. 
Figure~\ref{fig:jaccard} shows the relation between Jaccard distance and
iteration count for the number of chain $10$. As we can see, for increasingly
larger iteration count, the Jaccard distance among the top-$k$ patterns from
different chains diminishes and for sufficiently large value, it converges to a
small value.

\section{Conclusion}\label{myconclusion}
In this paper, we present \alg, a sampling based method for finding frequent
induced subgraph of a given size. For large input graphs, existing algorithms
for frequent subgraph mining are completely infeasible; whereas \alg\ can return
a small set of probabilistically frequent patterns of desired size within a
small amount of time. Our experiments show that the expected support of the
graphs that \alg\ samples has excellent rank correlation with their actual
support. The theoretical proof relevant in this paper has been presented in the appendix section.

\bibliographystyle{IEEEtran}
\bibliography{fscube}

\begin{appendix}
\setcounter{lemma} {0}
\begin{lemma} \label{score-lemma}
$s_1(g) \ge support(g)$ and $s_2(g) \ge support(g)$
\end{lemma}

\begin{proof}
Consider an edge $e \in E(g)$. Since $e \in E(g)$, 
$support\mbox{-}set(e) \supseteq support\mbox{-}set(g)$, hence $support(e) \ge support(g)$.
Since this hold for all the edges, average-support of the edges is an upper bound
of the support of $g$; hence, $s_1(g) \ge support(g)$. 

To compute $s_2(g)$ we intersect the $support\mbox{-}set(e)$ of all edges $e \in g$.
Thus, $s_2(g)$ considers the support of the edge-set of $g$, without considering
the graphical constraint imposed by $g$, so $s_2(g) \ge support(g)$.
\end{proof}

\begin{lemma} \alg's random walk is ergodic.
\end{lemma}

\begin{proof}
A Markov chain is ergodic if it converges to a stationary distribution. To
obtain a stationary distribution the random walk needs to be finite,
irreducible and aperiodic. The state space consisting of all $p$-subgraphs is finite for
a given $p$. We also assume that the input graph $G$ is connected, so in this random walk
any state $y$ is reachable from any state $x$ with a positive probability
and vice versa, so the random walk is irreducible. Finally, the walk can be
made aperiodic by allocating a self-loop probability at every node. Thus the
lemma is proved.
\end{proof}

\begin{lemma}
The random walk of \alg\ achieves the target probability distribution, which is proportional
to the chosen scoring function ($s_i$)
\end{lemma}

\begin{proof}
An ergodic random walk achieves the target probability distribution if it
satisfies the reversibility condition i.e., for two neighboring states $x$ and
$y$, $\pi(x)T(x,y) = \pi(y)T(y,x)$, where $\pi$ is the target distribution and
$T(x,y)$ is the transition probability from $x$ to $y$.  For \alg\ the target
distribution for a graph $x$, $\pi(x)= \frac{s_i(x)}{K}$, where $K$ is a
normalizing constant. Now, from Figure~\ref{fig:sub}, it is easy to see that
$\pi(x)T(x, y)= \frac{s_i(x)}{K \cdot d_x} \min\left\{1,\frac{d_x *
s_i(y)}{d_y*s_i(x)}\right\} = \frac{1}{K}
\min\left\{\frac{s_i(x)}{d_x},\frac{s_i(y)}{d_y}\right\}$. Since the neighborhood
relation is symmetric, there can be a transition from the state $y$ to $x$ and
using that we have $\pi(y)T(y, x) = \frac{s_i(y)}{K \cdot d_y}
\min\left\{1,\frac{d_y * s_i(x)}{d_x * s_i(y)}\right\} = \frac{1}{K}
\min\left\{\frac{s_i(y)}{d_y},\frac{s_i(x)}{d_x}\right\}$. So, $\pi(x)T(x,y) =
\pi(y)T(y,x)$, which proves the lemma.\end{proof}

{\bf Theoretical Analysis of \alg} \alg\ ranks the subgraph patterns based on the expected support ($support_a$).
In this section, we analyze the expected value of $support_a$ for a
$p$-subgraph pattern $g$.  To simplify the analysis, we will assume that in
each sampling iteration (in Line 5 of Figure~\ref{fig:alg}), \alg\ returns one
of the $p$-subgraphs of the chosen database graph uniformly. This assumption
actually perform a worst-case analysis, because in general \alg\ performs a
biased sampling in which the presumable frequent $p$-subgraphs are sampled with
higher probability.

Let, ${\cal G} =\{G_1, G_2, \ldots, G_n \}$ be a graph database with $n$
graphs. Let's use $x_j$ to denote the number of distinct
$p$-subgraphs in the graph $G_j$. Assume that the (induced) support of a
subgraph pattern $g$ in the database ${\cal G}$ is $s$, and the id of the
graphs in which $g$ occurs are $z_1, z_2, \cdots, z_s$.

If \alg\ makes $t$ sampling iterations, on average $t/n$ samples are obtained
from the graph $G_{z_i:1\le i \le s}$. Under the uniform sampling assumption,
the probability of sampling $g$ from $G_{z_i}$ in at least one of $t/n$
iterations is equal to $1-\left(1 - 1/x_{z_i}\right)^{t/n}$. Since the number
of sampling iterations is typically very large, the above term is equal to
$1-(1- \frac{t}{n\cdot x_{z_i}}) = \frac{t}{n\cdot x_{z_i}}$.  So, the expected
support of $g$, $E\big[support_a(g)\big]=\frac{t}{n} \times (1/x_{z_1} +
1/x_{z_2} + \cdots + 1/x_{z_s})$. If the number of samples are in the same order
as the number of $p$-subgraphs in the database graphs, the expected support
converges to the actual support and the estimation is unbiased. Note that,
even if the value of $x_{z_i}$ are large (in the order of millions), \alg\
can sample millions of iterations in a few minutes, thus it can bring the $support_a$
value close to the actual support effectively. On the other hand, existing
methods are not scalable as performing millions of SI test will take months, if not years.

However, \alg\ performs much better than a uniform sampler, as it actually performs a 
support-biased sampling. In real-life datasets, the support of $p$-subgraphs follows
a heavy-tail distribution, in which a small number of truly frequent patterns have
high support, but the majority of the $p$-subgraphs have small support. Thus, the
success probability of sampling a frequent pattern $g$ from the graph $G_{z_i}$ is 
much higher than $\frac{t}{n\cdot x_{z_i}}$. In Section~\ref{sec:impact}, we will
compare between \alg\ and a modified version of \alg\ that uses the uniform sampling 
to show that \alg's performance is substantially better.

\end{appendix}
\end{document}